\documentclass[12pt]{article}
\usepackage[english]{babel}
\usepackage[pdftex]{color}
\usepackage{amsmath}
\usepackage{graphicx}
\usepackage{hyperref}
\usepackage{amsmath}
\usepackage{amsfonts}
\usepackage{amssymb}

\begin{document}

\title{ \textbf{{}Gauge-invariant Lagrangians for  mixed-antisymmetric higher spin fields\thanks{Talk presented at SQS'15,  03 August – 08 August, 2015, at JINR, Dubna, Russia}}}
\author{\textsc{Alexander A. Reshetnyak\thanks{%
reshet@ispms.tsc.ru}} \\
Laboratory of Computer-Aided Design of Materials, Institute of \\
Strength Physics and Materials Science, 634021 Tomsk, Russia}
\date{}
\maketitle

\begin{abstract}

Lagrangian  descriptions of  irreducible and reducible integer higher-spin representations of the Poincare group
subject to a Young tableaux $Y[\hat{s}_1,\hat{s}_2]$  with two columns are constructed within a metric-like formulation in
a $d$-dimensional flat space-time on the basis of a BRST approach extending the results of  [arXiv:1412.0200[hep-th]].  A Lorentz-invariant resolution of the BRST complex within both the constrained and unconstrained BRST formulations produces a gauge-invariant Lagrangian entirely in terms of the initial tensor field $\Phi_{[\mu]_{\hat{s}_1}, [\mu]_{\hat{s}_2}}$ subject to $Y[\hat{s}_1,\hat{s}_2]$ with an additional tower of gauge parameters realizing the $(\hat{s}_1-1)$-th stage of reducibility with a specific dependence on the value $(\hat{s}_1-\hat{s}_2)=0,1,...,\hat{s}_1$. Minimal BRST--BV action is suggested, being proper solution to the master equation in the minimal sector and providing objects appropriate to construct interacting Lagrangian formulations with mixed-antisymmetric fields in a general framework.
\end{abstract}

\section{Introduction}

Some topical problems of high-energy physics are related to higher-spin (HS) field theory,
being part of the LHC experiment program. The so-called tensionless limit of (super)string theory \cite{tensionlessl}, operating an infinite tower of HS fields with integer and half-integer generalized spins,
incorporates HS field theory into superstring theory and turns it into a method of studying the classical
and quantum structure of the latter (for the current status of HS field theory, see the reviews \cite{reviews}, \cite{reviewsV},   \cite{reviews3}). This paper examines the construction of Lagrangian formulations (LFs) and
so-called BRST--BV master actions in the minimal sector of the field-antifield formalism for free integer
mixed-antisymmetric (MAS) tensor HS fields in a flat $\mathbb{R}^{1,d-1}$-space-time subject to an arbitrary Young
tableaux (YT) with $2$ columns, $Y[\hat{s}_1,\hat{s}_2]$, for $\hat{s}_1 \geq \hat{s}_2 $, in a metric-like
formalism on a basis of the BRST--BFV approach \cite{BFV}, \cite{BFV1}.

Irreducible Poincare or (anti)-de-Sitter ((A)dS) group representations in constant curvature space-times may
be described by mixed-symmetric (MS) HS fields subject to an arbitrary YT with $k$ rows,
$Y({s}_1,...,{s}_k)$ (the case of a symmetric basis), determined by more than one spin-like parameters ${s}_i$
\cite{Labastida},  \cite{metsaevmixirrep}, and, equivalently, by MAS (spin-)tensor fields subject to an arbitrary YT,
albeit with $l$ columns, $Y[\hat{s}_1,...,\hat{s}_l]$  (the case of an antisymmetric basis), with integers \cite{Reshetnyak_mas}, \cite{BurdikReshetnyak}  or half-integers $\hat{s}_1\geq \hat{s}_2\geq ...\geq \hat{s}_l$  having a spin-like interpretation.

MS and MAS HS fields arise in $d>4$ space-time dimensions in addition to totally symmetric
and antisymmetric irreducible representations of the Poincare and (A)dS algebras. For these latter,
as well as for MS HS fields, LFs for massless and massive free higher-spin fields are well-developed  (see, e.g. refs. in \cite{Reshetnyak_mas})
\cite{massless Minkowski},  \cite{massless AdS}, \cite{massive AdS},   \cite{SkvortsovZinoviev},
\cite{Zinovievta}, including the BRST--BFV approach, e.g., in \cite{BurdikPashnev}--\cite{Reshetnyk2}.
For MAS, the problem of field-theoretic description has not been solved completely, except for: massless constrained MS and MAS tensor fields on Minkowsky space, $\mathbb{R}^{1.d-1}$, as the elements of irreducible representations of $gl(d)$-algebra, in terms of  multiforms \cite{Latini1}  on a base of BRST detour quantization techniques \cite{Latini2}, with  Einstein operator for the equations of motion; constrained
bosonic MAS fields with 2 group indices:  Lagrangians for massless HS fields on Minkowsky and for  massive  HS fields on AdS spaces have been  considered respectively in \cite{MedeirosHall}, \cite{massive_AdS}, whereas the  massless fields  at the level of the equations of motion in a frame-like formulation were studied in
\cite{Alkalaev}.

The paper is organized as follows. In Section~\ref{BRSTBFV}, we remind the
key points of finding BRST--BFV Lagrangian formulations for MAS HS fields. In Section~\ref{metric}
we find gauge-invariant Lagrangians in explicit tensor forms on a basis of the BRST complex resolution.
The construction of a minimal master action and the possibility to deform its structure by non-quadratic
interacting terms with appropriate HS fields in the BRST--BFV approach are briefly examined
in Section~\ref{minimalBRSTBV}.

We use the convention $\eta_{\mu\nu} = diag (+,
-,...,-)$ for the metric tensor, with the Lorentz indices $\mu, \nu = 0,1,...,d-1$, and the notation $\epsilon(A)$, $[gh_{H},gh_{L}, gh_{tot}](A)$ for the respective values of Grassmann parity, BFV, $gh_{H}$, BV, $gh_{L}$ and total, $gh_{tot}=gh_{H}+gh_{L}$, ghost numbers of a quantity $A$.  The supercommutator $[A,\,B\}$ of quantities $A, B$
with definite values of Grassmann parity is given by $[A\,,B\}$ = $AB -(-1)^{\epsilon(A)\epsilon(B)}BA$.


\section{BRST-BFV Lagrangian formulations}

\label{BRSTBFV} 

Recall \cite{Reshetnyak_mas} that a massless integer-spin irreducible
representation of the Poincare group in Minkowski space
$\mathbb{R}^{1,d-1}$  is described by a rank-$(\hat{s}_1\hspace{-0.1em}+\hspace{-0.1em}\hat{s}_2)$ tensor field
$\Phi_{[\mu^1]_{\hat{s}_1},[\mu^2]_{\hat{s}_2}}
\hspace{-0.2em}\equiv \hspace{-0.2em}
\Phi_{\mu^1_1...\mu^1_{\hat{s}_1},\mu^2_1...\mu^2_{\hat{s}_2}}$
with generalized spin
 $\mathbf{s} \equiv (s_1,...,s_{s_2}; s_{s_2+1},...,s_{s_1})$ = $(2, 2,
 ... , 2; 1, ..., 1)$ (omitting the symbol "$\hat{\phantom{s}}$" under $\hat{s}_i$ and $s_1 \geq s_2 > 0, s_1
\leq [d/2]$), subject to a YT, $Y[s_1,s_2]$ with $2$ columns
of height  $s_1, s_2$.
The field satisfies differential equations (Klein--Gordon and divergentless ones)
(\ref{mEq-1}) and algebraic equations (traceless and mixed-antisymmetry ones) (\ref{mEq-3}) :
\begin{eqnarray}
\label{Eq-0b} &&\hspace{-1.5em}
\partial^\mu\partial_\mu\Phi_{[\mu^1]_{s_1},[\mu^2]_{s_2}}
 =0, \quad
 \partial^{\mu^i_{l_i}}\Phi_{
[\mu^1]_{s_1},[\mu^2]_{s_2}} =0, \texttt{ for } 1 \leq l_i \leq s_i,\ i=1,2,\label{mEq-1}
\\
&& \hspace{-1.5em}   \eta^{\mu^1_{l_1}\mu^2_{l_2}}\Phi_{
[\mu^1]_{s_1},[\mu^2]_{s_2}}=0,\texttt{ for }
 1 \leq l_i \leq s_i,  \quad \Phi_{
[[\mu^1]_{s_1},\underbrace{\mu^2_1...\mu^2_{l_2-1}}\mu^2_{l_2}]...\mu^2_{s_2}}=0, \label{mEq-3}
\end{eqnarray}
where the bracket  means that the indices inside do not take part in antisymmetrization.

Equivalently, the relations
\begin{eqnarray}
\label{equivH} \hspace{-1ex}&& \hspace{-2ex} \big({{l}}_0, {l}_i, l_{12},
t_{12} \big)|\Phi\rangle  =  \big(\partial^\mu\partial_\mu, -i {a}^i_\mu \partial^\mu, \textstyle\frac{1}{2}{a}^{1}_\mu {a}^{2\mu}, {a}^{1+}_\mu
a^{2\mu}\big) |\Phi\rangle = 0 , \\
\label{PhysState}  \hspace{-1ex}&& \hspace{-2ex} \mathrm{for} \ |\Phi\rangle  =
\sum_{s_1=0}^{[d/2]}\sum_{s_2=0}^{s_1}
\frac{\imath^{s_1+s_2}}{{s_1!s_2!}}\Phi_{[\mu^1]_{s_1},[\mu^2]_{s_2}}\,
\prod_{i=1}^2\prod_{l_i=1}^{s_i} a^{+\mu^i_{l_i}}_i|0\rangle,
\end{eqnarray}
describe all the integer spin MAS $ISO(1,d-1)$ group irreps with the help of a string-like vector
$|\Phi\rangle \in \mathcal{H}^f$ in an auxiliary Fock space
$\mathcal{H}^f$,  generated by $2$ pairs of fermionic oscillators
$a^i_{\mu^i}(x), a^{j+}_{\nu^j}(x)$: $\{a^i_{\mu^i},
a_{\nu^j}^{j+}\}=-\eta_{\mu^i\nu^j}\delta^{ij}$.
To describe the single Poincare group irrep of spin $\mathbf{s}=[s_1,s_2]$,
we extend (\ref{equivH}) by spin relations with the number particle operators $g_0^i$:
\begin{eqnarray}\label{mg0i}
g_0^i|\Phi\rangle =(s_i-\textstyle\frac{d}{2}) |\Phi\rangle, \ \mathrm{for} \
 g_0^i = -\frac{1}{2}[{a}^{i+}_\mu,  {a}^{\mu{}i}].
\end{eqnarray}
The condition that the BRST operator be hermitian leads to extending
the set of primary constraints $\{o_\alpha\}$
= $\bigl\{{{l}}_0, {l}_i, l_{12}, t_{12} \bigr\}$
by their hermitian conjugates with respect to the scalar product in $\mathcal{H}^f$:
\begin{eqnarray}
\label{fsprod} \hspace{-1ex} && \hspace{-2ex} \langle{\Psi}|\Phi\rangle  =   \int
d^dx\sum_{s_1=0}^{[d/2]}\sum_{s_2=0}^{s_1} \frac{(-1)^{s_1+s_2}}{s_1!s_2!}\Psi^*_{[\mu^1]_{s_1},[\mu^2]_{s_2}}(x)
\Phi^{[\mu^1]_{s_1},[\mu^2]_{s_2}}(x),
\end{eqnarray}
so that the total set of constraints together with $g_0^i$,
$o_I=\bigl\{{l}_0, {l}_{i}, l_{12}, t_{12}, {l}^+_{i}, l^+_{12}, t^+_{12} , g_0^i\bigr\}$,
forms a Lie superalgebra, $\mathcal{A}(Y[2],\mathbb{R}^{1,d-1})$.
The BRST operator $Q'$ which encodes $\mathcal{A}(Y[2],
\mathbb{R}^{1,d-1})$ has the same form as the one with the deformed superalgebra $\mathcal{A}_c(Y[2],
\mathbb{R}^{1,d-1})$ for the converted operators $O_I = o_I+o'_I$, obtained by resolving
the problem of an additive conversion for the subalgebra $sl(2)\oplus sl(2)$ in $\mathcal{A}(Y[2],
\mathbb{R}^{1,d-1})$, with constraints being only the algebraic second-class ones,
$l_{12}, t_{12}, l^+_{12}, t^+_{12}$, and the pair  $g_0^i$, which reflects this fact: $[l^+_{12},l_{12}\}=\frac{1}{4}\sum_ig_0^i$; $[ t^+_{12},  t_{12}\}=\sum_i(-1)^ig_0^i$. From the explicit form of Verma module
construction for the $sl(2)\oplus sl(2)$ algebra, and from presenting its elements $o'_I$
as polynomials in the new Fock space $H^{\prime}$ with 2 pairs of bosonic oscillators $b_i, b_i^+$ (for  $[b_i, b_j^+]=\delta_{ij}$), we present $o'_I$ \cite{Reshetnyak_mas} as follows:
\begin{eqnarray}
\Big[ g_{0}^{i\prime}, l_{12}', t_{12}'\Big]  =  \Big[  h_i +b_1^{+} b_1 +(-1)^i b_2^{+} b_2,-\textstyle \frac{1}{4} (\sum h_k + b_1^{+} b_1 ) b_1,   -( \sum(-1)^k h_k + b_2^{+} b_2 ) b_2\Big]\label{osc-}
\end{eqnarray}
and $\big[t^{+\prime}_{12}, {l_{12}^{+}}' \big] = \big[b_2^{+},  b_1^{+}\big]$ with yet undetermined $\mathbb{R}$-constants $h_i$.
In order to be hermitian-conjugated to each other for the corresponding pair, $t^{+\prime}_{12}$ and $t^{\prime}_{12}$, $l^{+\prime}_{12}$ and  $l^{\prime}_{12}$,  as well as to be hermitian self-conjugated for  $g_{0}^{i\prime}$, the Grassmann-even operator  $(K')^+=K'$ is introduced:
\begin{eqnarray}
\label{Chn}
 && \hspace{-2em} K' =  \sum_{n_{i} = 0}^{\infty}
 \frac{(-1)^{n_1+n_2}C_{h_1+h_2}(n_1)C_{h_2-h_1}(n_2)}{4^{n_1}n_{1}!n_{2}!(h_1+h_2+n_1)(h_2-h_1+n_2) }|n_1,n_2\rangle \langle n_1,n_2 |,\texttt{ for  } C_{h}(n) = \prod_{i=0}^n (h+i),\
\end{eqnarray}
with $|n_1,n_2\rangle  =  \prod_{i}(b_i^+)^{n_i}|0\rangle$ providing modified hermiticity properties in $\mathcal{H}'$:
\begin{eqnarray}
 (s_{12}')^+  K'\ =\  K' s_{12}^{\prime+}, \ \ (g_0^{i\prime})^+  K'\ =\  K' g_{0}^{i\prime}  \ \mathrm{for} \ s\in\{t,l\} .\label{systemK}
\end{eqnarray}
The hermitian nilpotent BRST operator $Q'$, $(Q')^2 = 0$,  for $\mathcal{A}_c(Y[2],
\mathbb{R}^{1,d-1})$ ($Q^{\prime +}K\  =\ K Q'$ with  $K = \ 1 \otimes K' \otimes 1_{gh_H}$ and $gh_H (Q')=1 $)   obtained with bosonic  $q_i^{(+)}$ and fermionic  $\eta_0, \eta_{12}^{(+)}, \vartheta_{12}^{(+)}, \eta^i_G$ coordinates  and corresponding  momenta  $p_i^{(+)}$ and  $\mathcal{P}_0, \mathcal{P}_{12}^{(+)},\lambda _{12}^{(+)}, \mathcal{P}^i_G$, having the opposite values of $gh_H$ for the respective elements $O_I$,  reads as follows:
\begin{eqnarray}\label{Q'}
\hspace{-1em}&&\hspace{-0.7em} Q'  =   Q +
\eta^i_{G}(\sigma^i+h^i)+\mathcal{D}^i  \mathcal{P}^i_{g}, \ \mathrm{with} \   Q = \eta_0 l_0\hspace{-0.1em} +\hspace{-0.1em} i  q_iq_i^+  P_0+\Delta Q,\\
 \hspace{-1em}&&\hspace{-0.7em}   \Delta Q  \hspace{-0.1em}=\hspace{-0.1em}  \Bigr( q_il^+_i \hspace{-0.1em} + \hspace{-0.1em} \eta_{12}L^+_{12} + \vartheta_{12} T^+_{12} + \textstyle\frac{1}{2}\epsilon_{ij}\eta_{12} q_i^+ p_j^+
  + \vartheta_{12}(q_2^+p_1+q_1p_2^+ ) +h.c.\Bigr)  \nonumber\\
\hspace{-1em}&&\hspace{-0.7em}
  \sigma_i+h_i=  G_0^{i} - q_i^+ p_i - q_i p_i^+   +\eta_{12}^+ \mathcal{P}_{12} -\eta_{12} \mathcal{P}^+_{12}  +(-1)^i( \vartheta_{12}^+\lambda_{12}-\vartheta_{12}\lambda_{12}^+ )
\label{separation2} ,
\end{eqnarray}
with $\epsilon_{ij}=-\epsilon_{ji}, \epsilon_{12}=1$, a generalized spin operator $\sigma_i$, and the only nontrivial (super) commutators  $[q_i, p^+_j]$= $-\imath\{\eta^i_G,\mathcal{P}^i_G,\}$=$\delta_{ij} $, $-\imath\{\eta_0,\mathcal{P}_0,\}$  =  $\{ \eta_{12},  \mathcal{P}_{12}^{+}\}$ = $ \{\vartheta_{12}, \lambda _{12}^+\}$ = $1$.

The Lagrangian formulation in the unconstrained case for an HS field with given spin $[s_1,s_2]=[s]_2$ is determined
by a gauge-invariant  action and a sequence of $(s_1 + s_2)$-stage reducible gauge transformations:
\begin{eqnarray} {\cal
S}_{[s]_2} = \int d \eta_0  {}_{[s]_2}\langle \chi^0 |K_{[s]_2}
Q_{[s]_2}| \chi^0 \rangle_{[s]_2},  \quad \delta|\chi^k \rangle_{[s]_2}
=Q_{[s]_2}|\chi^{(k+1)}\rangle_{[s]_2}, \mathrm{for }\ k=0,...,\sum_is_i
, \label{S}
\end{eqnarray}
with a non-gauge
$|\chi^{(\sum_is_i+1)}\rangle_{[s]_2}$, where $| \chi^0 \rangle_{[s]_2}\big|_{q^+,p^+...=0}= | \Phi \rangle_{[s]_2}$
(\ref{PhysState}), and with the substitutions $(K,Q)\big|_{h _i\to h_i(s) } = \big(K_{[s]_2},Q_{[s]_2}\big)$ made
for $h_i(s)$, being (together with field and gauge parameter vectors  $| \chi^k \rangle_{[s]_2}$, $k=0,1...$) eigen-values and eigen-vectors  for the spectral problem which follows from the BRST complex: $Q'|\chi^0\rangle=0$, $\delta|\chi^k\rangle=Q'|\chi^{k+1}\rangle$, $gh_H(|\chi^k\rangle)= -k$ in the representation $(q_i, p_i,  \eta_{12}$, $\vartheta_{12},
\mathcal{P}_0$, $ \mathcal{P}_{12}$, $\lambda_{12},
|0\rangle=0$ ;  $\mathcal{P}^{i}_G)|\chi^k\rangle =0$:
\begin{eqnarray}
\label{Qchi} \hspace{-1em}&&\hspace{-0.5em}  \bigl(Q|\chi^0\rangle, \delta|\chi^k\rangle\bigr) =\bigl(0, Q|\chi^{k+1}\rangle(1-\delta_{k,\sum_is_i+1})  \ \mathrm{and} \  [\sigma^i+h^i]|\chi^k\rangle = 0,  \\
\hspace{-1em}&&\hspace{-0.7em} |\chi^k\rangle =\hspace{-0.1em}
\sum_{\{n\}_b=0}^{\infty}
\hspace{-0.15em}\sum_{\{n\}_f=0}^{1}\hspace{-0.1em}
\eta_{0}^{n_{\eta_{0}}}
\eta_{12}^{+n_{\eta_{12}}}
\vartheta_{12}^{+n_{\vartheta_{12}}}
\mathcal{P}_{12}^{+n_{P_{12}}}
\lambda_{12}^{+n_{\lambda_{12}}} \prod_{i=1}^{2}
q_i^{+n_{q_i}}
p_i^{+n_{p_i}}
b_i^{+n_{b_i}}
\left|\Phi^k({a_i}^+)_{\{n\}_f \{n\}_b}\rangle\right.,
\label{extState}
\\
\label{hi} \hspace{-1em}&&\hspace{-1.5em} \ \mathrm{therefore} \   -h^i(s) = s_i - \frac{d+1-\delta(m,0)}{2}  -(-1)^i \;, \
 s_1, \in \mathbb{Z}, s_2 \in
\mathbb{N}_0\,
\end{eqnarray}
in the massless and massive cases, with  2 pairs of additional odd oscillators in $(L_i,L_i^+)=(l_i+mf_i,l_i^++mf_i^+)$.

In the constrained case, the only differential constraints compose a BRST complex, including a restricted BRST operator and a spin operator $(Q_r, \sigma^i_r)$ = $(Q, \sigma^i) \big|_{(\eta^{(+)}_{12}=\vartheta^{(+)}_{12}=0)}$, without conversion and with off-shell BRST-extended traceless and MAS constraints, which enter into
the definition of a constrained LF of $s_1+s_2-1$-stage of reducibility,
\begin{eqnarray}
\hspace{-0.0em}\mathcal{S}_{r[s]_2}\hspace{-0.1em} = \hspace{-0.1em}\int\hspace{-0.1em} d \eta_0  {}_{[s]_2}\langle \chi^0_r
|Q_r| \chi^0_r \rangle_{[s]_2}, \;  \Big(\delta;  {l}_{12}+  \textstyle\frac{1}{2}\epsilon_{ij} q_i p_j;   t_{12} +   q_2p_1^++q_1^+p_2\Big)| \chi^k_r \rangle =
\Big(Q_r| \chi^{k+1}_r \rangle;\vec{0}\Big) \label{Sr}
\end{eqnarray}
for $k=0,1,...,s_1+s_2$.  In the $\eta_0$-independent form, $| \chi^0_r \rangle=| S^0_r \rangle+ \eta_0| B^0_r \rangle$, the action reads
\begin{eqnarray}&& \mathcal{S}_{r[s]_2} =  (-1)^{s_1+s_2} {}_{[s]_2}\Big(\langle S^{0}_{r}\big| ,\langle B^{0}_{r}\big|\Big)\left(\begin{array}{cc}
                                                                                          l_0 & - \Delta Q_r  \\
                                                                                          -\Delta Q_r  &  q_iq_i^+
                                                                                        \end{array}\right)\left(\begin{array}{c}
                                                                                          \big|S^{0}_{r} \rangle_{[s]_2} \\
                                                                                          \big|B^{0}_{r} \rangle_{[s]_2}
                                                                                        \end{array}\right)
 .\label{Sclfvectf}
\end{eqnarray}
Note, without off-shell constraints, we have obtained from (\ref{Sr}) a generalized triplet formulation, which describes reducible Poincare group representations with different spins.

\section{Metric-like  component Lagrangians from\hspace{-0.1em} BRST \hspace{-0.2em} com\-plex resolution}

\label{metric}
The deduction of component Lagrangians (for $m=0$) entirely in terms of the initial HS tensor field
$\Phi_{[\mu^1]_{{s}_1},[\mu^2]_{{s}_2}}$ is based on a partial gauge-fixing procedure for unconstrained
LF (see \cite{BurdikReshetnyak} for details):
\begin{enumerate}
    \item removing the dependence on the  auxiliary oscillators $b_i^+, i=1,2$ in $| \chi^{k} \rangle_{[s]_2}$ by means of the gauge transformations (\ref{Qchi}) resulting in the gauge conditions: $
\Big(b_{2}\lambda_{21}^+$, $b_{1}\mathcal{P}_{12}^+\lambda_{21}^+\Big)|\chi^{k}\rangle$ $=0$,  for  $k=0,\ldots, \sum_{i}^2 s_i$;
      \item  removing the dependence on the ghosts $\vartheta_{12}^+, \eta_{12}^+$  in $| \chi^{k} \rangle_{[s]_2}$ via the residual gauge transformations (\ref{Qchi}) and the equations of motion, which  leads to the vanishing of  $|\chi^{l}\rangle=0$, $l= s_1+1,...\sum_i s_i$ and to algebraic relations on the final field and gauge vectors, so that  the surviving $|\chi^k\rangle=0$, $k=0,...,s_1$ depend entirely on the single (restricted only by partial mixed symmetry)
component tensor field.
\end{enumerate}
Calculating the scalar products for the resulting Lagrangian action \begin{eqnarray}
&\hspace{-0.5em}& \hspace{-2.0em} \mathcal{S}_{[s]_2} \hspace{-0.1em}= \hspace{-0.1em} (-1)^{\sum s_i}\hspace{-0.15em}{}_{[s]_2}\hspace{-0.1em}\langle \Phi\big|\hspace{-0.1em}\bigg\{\hspace{-0.1em}\sum_{r=0}^{s_2}\frac{(-1)^{r}2^{2r}}{r!(r+1)!} (l_{12}^+)^r  \hspace{-0.1em}\Big[l_0   - \left\{1+\textstyle\frac{r}{2}\right\}\sum_il_i^+ l_i      -\frac{2}{r+2} l_{12}^+\sum_il_i l_i^+l_{12} \nonumber \\
&& \phantom{ \mathcal{S}_{[s]_2}  =}
+ 2l_{12}^+l_2l_1+  2l_1^+l_2^+l_{12}\Big]\hspace{-0.1em}  l_{12}^r\hspace{-0.1em}\bigg\}\hspace{-0.1em} \big|\Phi\rangle_{[s]_2}\label{Sclkin}
\end{eqnarray}
and expressing the tensor components in the remaining tower of gauge transformations leads to the final LF for the field  $\Phi_{[\mu^1]_{{s}_1},[\mu^2]_{{s}_2}} $ subject to $Y[s_1,s_2]$ of spin $[s]_2$:
 \begin{eqnarray}
&& \hspace{-1em} \mathcal{S}_{[s_2]}(\Phi) \hspace{-0.1em} = \hspace{-0.2em} \int \hspace{-0.2em}d^dx \hspace{-0.1em}\bigg\{\hspace{-0.15em}\sum_{r=0}^{s_2}\frac{(-1)^{r}}{(s_1-r)!(s_2-r)!r!(r+1)!}
(\mathrm{Tr}^r\Phi)_{[\mu^1]_{s_1-r},[\mu^2]_{s_2-r}} \hspace{-0.1em}\Big[
\partial^2  (\mathrm{Tr}^r\Phi)^{[\mu^1]_{s_1-r},[\mu^2]_{s_2-r}}
   \nonumber \\
 &&  - \left\{1+\textstyle\frac{r}{2}\right\} \Big( (s_1-r){\partial}^{\mu^1_1}   {\partial}_{\nu^1_1}\delta_{\nu^2_1}^{\mu^2_1}+ (s_2-r){\partial}^{\mu^2_1}   {\partial}_{\nu^2_1} \delta_{\nu^1_1}^{\mu^1_1}\Big) (\mathrm{Tr}^r\Phi)^{\nu^1_1\mu^1_2...\mu^1_{s_1-r}, \nu^2_1\mu^2_2...\mu^2_{s_2-r}}    \nonumber \\
 &&
     -\frac{(s_1-r)(s_2-r)}{2(r+2)}  \eta^{\mu^1_{s_1-r}\mu^2_1}  \Big\{2\partial^2(\mathrm{Tr}^{r+1}\Phi)^{[\mu^1]_{s_1-r-1},\mu^2_2...\mu^2_{s_2-r}}
      - \Big( (s_2-r-1) {\partial}^{\mu^2_2}   {\partial}_{\rho^2_1}\delta_{\rho^1_1}^{\mu^1_1}\nonumber \\
 && \ +(s_1-r-1) {\partial}^{\mu^1_2}   {\partial}_{\rho^2_1}\delta_{\mu^2_2}^{\mu^1_1}\Big)(\mathrm{Tr}^{r+1}\Phi)^{\rho^1_1\mu^1_2...\mu^1_{s_1-r-2},\rho^2_1\mu^2_3...\mu^2_{s_2-r}}\Big\}
      \nonumber \\
 &&  + \frac{1}{2} (s_1-r)(s_2-r)\Big\{\eta^{\mu^1_{s_1-r}\mu^2_1} {\partial}_{\mu^1_{s_1-r}}   {\partial}_{\mu^2_1}(\mathrm{Tr}^{r}\Phi)^{[\mu^1]_{s_1-r},[\mu^2]_{s_2-r}} \nonumber \\
 &&
+  \eta_{\rho^1_{s_1-r}\rho^2_1} {\partial}^{\mu^1_{s_1-r}}   {\partial}^{\mu^2_1}(\mathrm{Tr}^{r}\Phi)^{[\mu^1]_{s_1-r-1}\rho^1_{s_1-r}, \rho^2_1\mu^2_2...\mu^2_{s_2-r}}\Big\}    \Big]\bigg\}.\label{Sclfcomp2}
\end{eqnarray}
with the notation for a multiple trace, $(\mathrm{Tr}^r \Phi)_{[\mu^1]_{s_1-r},[\mu^2]_{s_2-r}}\equiv \Phi_{[\mu^1]_{s_1-r}\nu_1...\nu_r,}{}^{\nu_r...\nu_1}{}_{[\mu^2]_{s_2-r}}\equiv \prod_{i=1}^r\eta^{\mu^1_{s_1+1-i} \nu^2_{i} }\Phi_{[\mu^1]_{s_1},[\mu^2]_{s_2}}$, and for the gauge-independent $s_1$-level gauge tensor parameter $\varphi^{s_1}_{[\mu^2]_{s_2}}$ in terms of $\big|\varphi^{s_1}(a^+)\rangle_{[0,s_2]}$  [with  $s_2=s_1-l$]:
\begin{eqnarray}
\label{gtrl=s1-1}  && \delta \big|\varphi^{s_1-k}(a^+)\rangle_{[k,s_2]} = -\Big((s_1-k+1) l_1^+ +   l_2^+t^{}_{12}  \Big) \big|  \varphi^{s_1-k+1}(a^+)\rangle_{[k-1,s_2]},   \\
\label{gtrl=s1-2}   && \ \mathrm{for} \left\{\begin{array}{l}  t_{12}^{s_2} \big|\varphi^{s_1-k}(a^+)\rangle_{[k,s_2]} \in Y[s_2+k,0],\ k=0,...,l; \\
t_{12}^{s_1-k} \big|\varphi^{s_1-k}(a^+)\rangle_{[k,s_2]} \in Y[s_1,s_2-s_1+k],\  k=l+1 ,...,s_1\end{array}\right.,
\end{eqnarray}
where the fact of belonging to the YT is understood for component tensor gauge parameters $\varphi^{s_1-k}_{[\mu^1]_{k},[\mu^2]_{s_2}}$ of rank $(k+s_2)$ in $\big|\varphi^{s_1-k}(a^+)\rangle_{[k,s_2]}$ (with the structure being as in
(\ref{PhysState})) after respective multiple application of the Young antisymmetrization realized by $t_{12}^{s_2}, t_{12}^{s_1-k}$.
In terms of the tensor relations  we represent (\ref{gtrl=s1-1}), (\ref{gtrl=s1-2}) as
 \begin{eqnarray}
\label{gtrl=s1-1c}  &&  \delta \varphi^{s_1-k}_{[\mu^1]_{k},[\mu^2]_{s_2}}  =  (s_1-k+1) \partial_{[\mu^1} \varphi^{s_1-k+1}_{[\mu^1]]_{k-1}[\mu^2]_{s_2}} + (-1)^k\partial_{[\mu^2_1} (Y\varphi^{s_1-k+1})_{[[\mu^1]_{k-1},{\mu^1}]\mu^2_2...\mu^2_{s_2}]}, \\
&&  \ \mathrm{for} \left\{\begin{array}{l}  (Y^{s_2}\varphi^{s_1-k}) _{[[\mu^1]_k,\mu^1_{k+1}...\mu^1_{k+s_2}]}  \in Y[s_2+k,0],\ k=0,...,l; \\
(Y^{s_1-k}\varphi^{s_1-k}) _{[[\mu^1]_k,{\mu^1_k+1}...\mu^1_{s_1}] [\mu^2]_{s_2-s_1+k}}   \in Y[s_1,s_2-s_1+k],\  k=l+1,...,s_1\end{array}\right..
\label{gtrl=s1-2c} \\
&& \ \mathrm{where} \  (Y\varphi^{s_1-k})_{[[\mu^1]_k,{\mu^1}]\mu^2_2...\mu^2_{s_2}}  =  -   s_2 \varphi^{s_1-k}_{[[\mu^1]_k,{\mu^1}]\mu^2_2...\mu^2_{s_2}},    \nonumber \end{eqnarray}
and antisymmetrization in $\partial_{[\mu^1} \varphi_{[\mu^1]]_{k-1}[\mu^2]_{s_2}}$, $(Y\varphi^{s_1-k})_{[[\mu^1]_k,{\mu^1}]\mu^2_2...\mu^2_{s_2}}$ ($\partial_{[\mu^2_1} (\varphi^{s_1-k})_{[\mu^1]_{k},\mu^2_2...\mu^2_{s_2}]}$) contains the factor $1/k$ ($1/s_2$)\footnote{The Lagrangain formulations for massless bosonic HS fields subject to $Y[s_1,s_2]$ \cite{MedeirosHall} does not contain the tower of reducible gauge transformations} .
The resulting LF is a gauge theory of $(s_1-1)$-th stage of reducibility, which describes the free dynamics of a massless Bose-particle of spin $[s_1,s_2]$, with the single off-shell restriction of Young symmetry on the field
$\Phi\equiv \varphi^{0}$ and the gauge parameters $\varphi^{1},..., \varphi^{s_1}$.
If we derive a component LF without auxiliary tensor fields, starting from the constrained BRST LF
(\ref{Sr}), we shall come to the same result (\ref{Sclfcomp2}), (\ref{gtrl=s1-1c}).

\section{Minimal BRST-BV actions and interacting problem}
\label{minimalBRSTBV}

For simplicity, we consider a component LF and introduce a total set of minimal ghosts and their antifields, according to the rule (for the vanishing value of $gh_H$):
 \begin{eqnarray}\label{gho}
   && \varphi^{s_1-k}_{[\mu^1]_{k}, [\mu^2]_{s_2}} \to C^{s_1-k}_{[\mu^1]_{k},[\mu^2]_{s_2}}:  (\varepsilon,  gh_L) C^{s_1-k}_{[\mu^1]_{k},[\mu^2]_{s_2}} = s_1-k , \ \ k=0,...,s_1,\\
   && C^{s_1-k}_{[\mu^1]_{k},[\mu^2]_{s_2}} \to   C^{s_1-k|*}_{[\mu^1]_{k},[\mu^2]_{s_2}}:  \big(\varepsilon,  -gh_L\big) C^{s_1-k|*} = 1+s_1-k,
 \end{eqnarray}
with an odd $\Phi^*_{[\mu^1]_{s_1}, [\mu^2]_{s_2}}\equiv   C^{0|*}_{[\mu^1]_{s_1},[\mu^2]_{s_2}}$,  $C^{0}\equiv \varphi^{0}=\Phi$, where $C^{s_1-k|*}, C^{s_1-k}$ are subject to the Young symmetry constraints (\ref{gtrl=s1-2c}).
The minimal BRST--BV action, $\mathcal{S}^{\min}_{[s_2]}=\mathcal{S}^{\min}_{[s_2]}\big( C^0,...,C^{s_1}, C^{0|*},..., C^{s_1|*}\big)$, for a free HS field
\begin{eqnarray} \label{minact}
&& \hspace{-1em}  \mathcal{S}^{\min}_{[s_2]}  = \mathcal{S}_{[s_2]} +
\int d^dx \sum_{k=0}^{s_1-1}C^{k|*}{}^{[\mu^1]_{s_1-k},[\mu^2]_{s_2}} \Big[(k+1) \partial_{[\mu^1} C^{k+1}_{[\mu^1]_{s_1-k-1}],[\mu^2]_{s_2}} \nonumber \\
 && \ + (-1)^{s_1-k}\partial_{[\mu^2_1} (YC^{k+1})_{[[\mu^1]_{s-k-1},{\mu^1}]\mu^2_2...\mu^2_{s_2}]}\Big]  \end{eqnarray}
being  a proper solution of a master equation in terms of an odd Poisson bracket in the field-antifield space $\mathcal{M}$:
\begin{eqnarray}\label{meq}
&&\hspace{-2em}   \big(\mathcal{S}^{\min}_{[s_2]}, \mathcal{S}^{\min}_{[s_2]}\big)_{[s]_2} = 2 \int d^dx \mathcal{S}^{\min}_{[s_2]}\left(\sum_{k=0}^{s_1}\frac{\overleftarrow{\delta}}{\delta C^{{k}|[\mu^1]_{s_1-k}, [\mu^2]_{s_2}}(x)}\frac{\overrightarrow{\delta}}{\delta C^{k|*}_{[\mu^1]_{s_1-k}, [\mu^2]_{s_2}}(x)} \right)\mathcal{S}^{\min}_{[s_2]} = 0, \\
&&\hspace{-2em} \mathrm{for} \, \big(F, G\big)_{[s]_2} \hspace{-0.15em}\stackrel{def}{=} \hspace{-0.15em}\int\hspace{-0.15em} d^dx F\hspace{-0.2em} \left(\hspace{-0.2em}\sum_{k=0}^{s_1}\frac{\overleftarrow{\delta}}{\delta C^{{k}|[\mu^1]_{s_1-k}, [\mu^2]_{s_2}}(x)}\frac{\overrightarrow{\delta}}{\delta C^{k|*}_{[\mu^1]_{s_1-k}, [\mu^2]_{s_2}}(x)} - \big(C^{k} \leftrightarrow C^{k|*}\big) \hspace{-0.2em}\right)G
\end{eqnarray}
given for any functionals $ F.G \in C[\mathcal{M}]$.
The action  serves to construct a quantum action under an appropriate choice of a tower of reducible gauge conditions (e.g. for the  MAS field $\Phi_{[\mu^1]_{s_1},[\mu^2]_{s_1}}$:
$(\partial^{\mu^i}, \mathrm{Tr})\Phi_{[\mu^1]_{s_1},[\mu^2]_{s_1}}=(0,0)$, so that the gauge-fixed Lagrangian should contain only term with $\partial^2$) in the extended field-antifield space, as well as to find an interacting theory, including the MAS HS field $\Phi^a_{[s]_2}$, $a=1,2,...$, with a vertex at least cubic in  $\Phi^a_{[s]_2}$\footnote{When the paper was published in the Arxiv, we have known on the results of no self-interaction vertexes \cite{Boulanger1, Boulanger2}  which could  extend  the Lagrangian formulation for free MAS HS field $\Phi^1_{[s]_2}\equiv \Phi_{[s]_2}$, but with tower of reducible gauge transformations (see Eq.(2.2)  in \cite{Boulanger2}) being differed with ones given by (\ref{gtrl=s1-1c}), (\ref{gtrl=s1-2c}). Note, the gauge transformations (2.2) for $s_1>s_2$  in \cite{Boulanger2}  and  (2.3), (2.12) for $s_1=s_2$  in \cite{Boulanger1} did not preserve the respective Young symmetry properties for the gauge transformed field and gauge parameters, e.g. $(\Phi_{[s]_2}+\delta \Phi_{[s]_2}) \not\in Y[s_1,s_2]$ for $\Phi_{[s]_2}\in Y[s_1,s_2]$  and so on.}, as well as $\Phi_{[s]_2}$ interacting  with an external electromagnetic field  and some other HS fields which realize another Poincare group irrep, e.g., a totally-symmetric HS field in the Fronsdal formulation, like cubic interaction with gravitational field for simple mixed-symmetric fields in the frame-like formulation \cite{Boulanger3}. The consistency of deformation is to be controlled by the master equation for the deformed action with the interaction terms, thus producing a sequence of relations for these terms.

Notice, first, that the LF in the case of massive MAS HS fields in $\mathbb{R}^{1,d-1}$ may be obtained by dimensional reduction of the massless theory in $\mathbb{R}^{1,d}$, leading to a non-gauge theory.  Second, the metric-like LF (\ref{Sclfcomp2}), (\ref{gtrl=s1-1c}) may be deformed to describe dynamic of  both MAS HS field with spin $[s]_2$  on the AdS(d) space and,  independently, dynamic of MAS conformal  HS field  on $\mathbb{R}^{1,d-1}$ which, in turn maybe used to study AdS/CFT correspondence problem. The Fock-space inspired Lagrangian quantization for considered BRST-BFV LFs may be described with help of generalized field-antifield vector, $\big|\chi^{0}_{g|(r)}\rangle_{[s]_2}$ containing in its decomposition in powers of ghosts whole set of ghost fields and theirs antifields as considered in \cite{Grigoriev} for totally-symmetric  and suggested in \cite{Grigoriev2} for mixed-symmetric bosonic constrained HS fields.

\vspace{-1ex}

\paragraph{Acknowledgements}

The author is grateful to the organizers of the International Workshop SQS'15 for their hospitality.
I also thank I.L. Buchbinder, Yu.M. Zinoviev,  K.V.~Ste\-panyantz, P.Yu. Moshin
for their interest, useful discussions, M. Grigoriev, K. Alkalaev,  E. Latini and  N. Boulanger for valuable correspondence. The study was fulfilled within the RFBR Project
No. 17-02-01333.

\end{document}